# Extending Regression Without Truth to Integrate Ground-Truth Measurements for Evaluating Quantitative Imaging Methods with Patient Data

Yan Liu[a], Abhinav K. Jha[a,b,*]

[a]Department of Biomedical Engineering, Washington University in St. Louis, St. Louis, MO, USA;
[b]Mallinckrodt Institute of Radiology, Washington University in St. Louis, St. Louis, MO, USA

## ABSTRACT

Objective evaluation of quantitative imaging (QI) methods with patient data is often hindered by the lack of gold standards. To address this challenge, a class of regression-without-truth (RWT) techniques have been developed. These techniques assume that the true and measured values are linearly related and estimate the linear-relationship parameters without access to true values. However, reliable estimation of these parameters typically requires many patient samples, which can be expensive and time consuming to obtain, and even impossible in settings such as studies with rare diseases or with new clinical imaging procedures. Thus, there is an important need for strategies to perform evaluation of quantitative imaging methods with a small number of patient samples. In this context, we note that datasets with known ground truth, such as physical phantom studies, could be available. In this manuscript, we propose an approach that integrates information from both patient data without ground truth and known-ground-truth datasets to perform objective evaluation of QI methods. We validated the proposed approach using numerical studies, which showed that the proposed approach yielded improved performance in ranking QI methods compared with RWT technique. The results demonstrate the potential of the proposed approach for evaluating QI methods when patient data are limited and motivate further validation with clinically realistic simulation studies and clinical data.

## 1. INTRODUCTION

Quantitative imaging (QI), the extraction and use of numerical measurements from medical images to help with clinical decision making[1,2], is emerging as an important tool across a wide range of applications. Examples include the use of myocardial blood flow from cardiac positron emission tomography (PET) to help diagnose coronary artery disease[3], quantifying radiotracer uptake using single-photon emission computed tomography (SPECT) for dosimetry in radiopharmaceutical therapy[4] and measuring apparent diffusion coefficient using diffusion magnetic resonance imaging (MRI) to monitor response to cancer treatment[5]. Given the importance of these clinical applications, different QI methods have been and are being actively developed[6–9]. For clinical translation of QI methods, objective evaluation of these methods on the task of reliably measuring the underlying true quantitative values is crucial. Performing such evaluation with patient data is highly desirable but often hindered by the lack of gold standards. Thus, there is a critical need for objective evaluation methods that can be applied in the absence of gold standards.

To address this need, a class of regression-without-truth (RWT) techniques has been developed[10–13]. RWT techniques are based on the premise that while the true quantitative values are unknown, the measured values are result of specific image-formation and quantification processes applied to the true values, and thus, the true and measured values are expected to be mathematically related. More specifically, the RWT techniques posit a linear stochastic relationship that is characterized by slope, bias, and noise standard deviation parameters, and these parameters are then estimated using a maximum-likelihood (ML)-based approach without access to the true values. The ratio of estimated noise standard deviation to slope, termed as noise-to-slope ratio (NSR), is computed as a figure of merit to rank the QI methods based on precision. The efficacy of the RWT techniques has been demonstrated in multiple applications including comparing software packages in SPECT[14] and segmentation methods in cardiac cine MRI[15] on the task of measuring cardiac ejection fraction, comparing segmentation methods in diffusion MRI for estimating apparent diffusion coefficient[16], evaluating reconstruction methods in SPECT for estimating mean regional uptake[12,17] and segmentation methods in PET for measur-ing metabolic tumor volume[18]. These applications demonstrate the promise of RWT to evaluate QI methods in the absence of ground truth.

Despite the promise of RWT techniques, a major challenge for broader applications of these techniques is the requirement for large patient cohorts[12]. This requirement arises because the RWT techniques involve estimation of many parameters. For example, when evaluating 3 QI methods, around 13 parameters need to be estimated, including 3 slopes, 3 biases, 3 noise standard deviations and 4 parameters characterizing the true-value distribution when a four-parameter beta distribution is assumed. Consequently, reliable parameter estimation may require measurements from a large patient cohort. Obtaining a large number of such patient studies can be expensive, and, in many cases, impossible. For example, in rare diseases, patient availability is inherently limited. Similarly, consider studies involving new contrast agents, such as new tracers in PET or new isotopes for theranostic applications, in which different methods are evaluated for estimating certain quantitative value, e.g. the absorbed dose in lesions and radiosensitive organs. For these new agents, early-phase clinical trials often involve very few patients. In these cases, identifying the most reliable quantitative imaging method becomes challenging with limited data. Thus, there is an important need for strategies to perform evaluation of quantitative imaging methods with a small number of patient samples. In this context, we note that data sources with known ground truth, such as physical phantom studies, are available. Such data could provide additional information about the linear-relationship parameters, which can potentially be incorporated to improve the estimation of these parameters[19]. In this manuscript, our objective is to extend the RWT technique to integrate such known-ground-truth data and investigate whether incorporating this additional information can lead to improved performance in ranking QI methods.

## 2. METHODS

### 2.1 Theory

Consider a scenario where $P$ patients are scanned by an imaging system. From the acquired data, $K$ QI methods are used to measure certain quantitative values. For the $p^{th}$ patient, the true quantitative value is denoted as $a_p$, the measured value yielded by the $k^{th}$ QI method is denoted as $\hat{a}_{p,k}$. Also, consider another study with $Q$ samples is conducted in a ground-truth-known setting, such as physical phantom study. The same $K$ QI methods are used to measure the quantitative values for these samples. For the $q^{th}$ physical phantom, denote the true quantitative value as $d_q$, the measured value yielded by the $k^{th}$ QI method as $\hat{d}_{q,k}$.

Similar as previous RWT technique[10–12], for the $k^{th}$ QI method, the measured values and true values are assumed to be linearly related by slope $u_k$, bias $v_k$ and a zero-mean Gaussian noise with standard deviation $\sigma_k$. Thus, for the $p^{th}$ patient, the relationship between true and measured values can be written as

$$\hat{a}_{p,k} = u_k a_p + v_k + \mathcal{N}(0, \sigma_k^2) \qquad (1)$$

For the $q^{th}$ physical phantom, the relationship between true and measured values can be written similarly as

$$\hat{d}_{q,k} = u_k d_q + v_k + \mathcal{N}(0, \sigma_k^2) \qquad (2)$$

For simplicity of notation, denote the measurements for the $p^{th}$ patient yielded by $K$ QI methods, $\{\hat{a}_{p,k}, k = 1, ..., K\}$, by $\widehat{\boldsymbol{A}}_p$, the measurements for the $q^{th}$ physical phantom, $\{\hat{d}_q^k, k = 1, ..., K\}$, by $\widehat{\boldsymbol{D}}_q$, the matrix containing slope and bias parameters, $\{u_k, v_k, k = 1, ..., K\}$, by $\boldsymbol{\Theta}$, the noise standard deviation parameters, $\{\sigma_k, k = 1, ..., K\}$, by $\boldsymbol{\Sigma}$. Based on Eq. (1)-(2), we can obtain the probability of observing $\widehat{\boldsymbol{A}}_p$ and the probability of observing $\widehat{\boldsymbol{D}}_q$, which both depend on the true value $a_p$ and true value $d_q$. For physical phantom study, we know the true value. However, for the patient data, the ground truth is unknown. To address this issue, we assume that the true values of the patient dataset are sampled from a parametric distribution characterized by $\boldsymbol{\Omega}$. We can then write the probability of observing the measurements of the $p^{th}$ patient, $\widehat{\boldsymbol{A}}_p$, without access to the true value as

$$\mathrm{pr}(\widehat{\boldsymbol{A}}_p | \boldsymbol{\Theta}, \boldsymbol{\Sigma}, \boldsymbol{\Omega}) = \int \mathrm{pr}(\widehat{\boldsymbol{A}}_p | a_p, \boldsymbol{\Theta}, \boldsymbol{\Sigma}) \mathrm{pr}(a_p | \boldsymbol{\Omega}) da_p . \qquad (3)$$

where $\mathrm{pr}(x)$ denotes the probability of a random variable $x$.

Denote all the measurements for $P$ patients, $\{\widehat{\boldsymbol{A}}_p, p = 1, ..., P\}$, as $\widehat{\boldsymbol{\mathcal{A}}}$, all the true values for $Q$ physical phantoms, $\{d_q, q = 1, ..., Q\}$, as $\boldsymbol{D}$ and all the corresponding measurements $\{\widehat{\boldsymbol{D}}_q, q = 1, ..., Q\}$, as $\widehat{\boldsymbol{\mathcal{D}}}$. Assuming independence among the true values of the *P* patients and given that the patient data and physical phantom data are independent, we can use an ML approach to estimate the values of $\{\Theta, \Sigma, \Omega\}$. The ML estimate of $\{\Theta, \Sigma, \Omega\}$ is given by

$$\{\Theta, \Sigma, \Omega\}_{ML} = \arg\max\{\log \text{pr}\left(\widehat{\mathcal{A}}|\Theta, \Sigma, \Omega\right) + \log \text{pr}\left(\widehat{\mathcal{D}}|\Theta, \Sigma, D\right)\}. \qquad (4)$$

Same as in the RWT technique[10,11], after the estimation of the linear-relationship parameters, the ratio of estimated noise standard deviation, $\sigma_k$, and slope, $u_k$, is used to quantify the precision of the $k^{\text{th}}$ method. This figure of merit, $\sigma_k/u_k$, termed as the noise-to-slope ratio, can thus be used to rank the QI methods. A lower value of NSR indicates a higher precision in measuring the true values.

## 2.2 Validating the proposed approach using numerical studies

We conducted numerical studies to investigate the impact of integrating known-ground-truth data using the proposed approach by comparing the RWT technique and the proposed approach in estimating NSR values and ranking QI methods. The numerical studies provided a controlled setting in which all model assumptions were satisfied and the true NSR values and true rankings of the QI methods were known.

Specifically, we first sampled *N* true values from a known four-parameter beta distribution. From these true values, we generated synthetic measurements for three hypothetical QI methods using specified slopes, biases, and noise standard deviations. This dataset was used to mimic patient data where the ground truth is unknown. We then input the noisy measurements into the RWT technique. For the proposed approach, we generated an additional dataset following the same procedure. This dataset was used to mimic a study in which ground truth is available, such as physical phantom study. The measurements from the ground-truth-unknown dataset and both the measurements and true values from ground-truth-known dataset were provided to the proposed approach.

To make the numerical studies clinically relevant, we used linear-relationship parameters derived from a realistic simulation study where different quantitative SPECT methods were used to quantify mean regional activity uptake in patients treated with $^{223}$Ra. Specifically, the slopes, biases, and noise standard deviations for the three synthetic QI methods were set to {0.97,0.76,1.00}, {0.30, 0,0} and {0.21,0.13,0.09}, respectively. The size of the ground-truth-unknown dataset was varied from 10 to 100. For the proposed approach, we additionally varied the number of known-ground-truth samples to assess how different amounts of such data affect performance. For each combination of patient cohort size and known-ground-truth dataset size, 200 noise realizations were repeated to estimate the percentage of correctly ranking all the QI methods and identifying the most precise QI method.

## 3. RESULTS

Figure 1 presents the average absolute normalized bias and average normalized standard deviation in estimating NSR values for RWT and the proposed approach with different numbers of samples in the known-ground-truth dataset. We observe that the proposed approach yielded more accurate and precise NSR estimation than RWT, indicating that integrating known-ground-truth data can improve estimation performance. Moreover, increasing the size of the known-ground-truth dataset further improved the accuracy and precision in estimating NSR. Consequently, the proposed approach achieved higher accuracy in correctly ranking the methods and identifying the most precise method compared with RWT, as shown in Figure 2. The ranking performance of the proposed approach improved consistently as the known-ground-truth dataset size increases.

## 4. DISSCUSSIONS AND CONCLUSION

One challenge in applying RWT techniques in clinical practice is the need for large patient datasets, which could be difficult to obtain in many clinical scenarios. To address this challenge, we investigated whether additional information from data sources with known ground truth, such as phantom studies or realistic simulation studies, could be integrated to improve performance of these RWT techniques. We derived a maximum likelihood approach that integrates both data with unknown ground truth (such as patient data) and data sources with known ground truth (such as physical phantom data). We validated the proposed approach using numerical experiments in this manuscript.

Our results from the numerical study showed that the proposed approach consistently yielded lower bias and lower standard deviation in estimating NSR compared with RWT. This then led to an improvement in the performance in ranking

QI methods. We note that the proposed approach achieved nearly 100% accuracy in identifying the most precise method using only 10 known-ground-truth samples and 10 patient samples (Figure 2(B)), where the ground truth is unavailable.

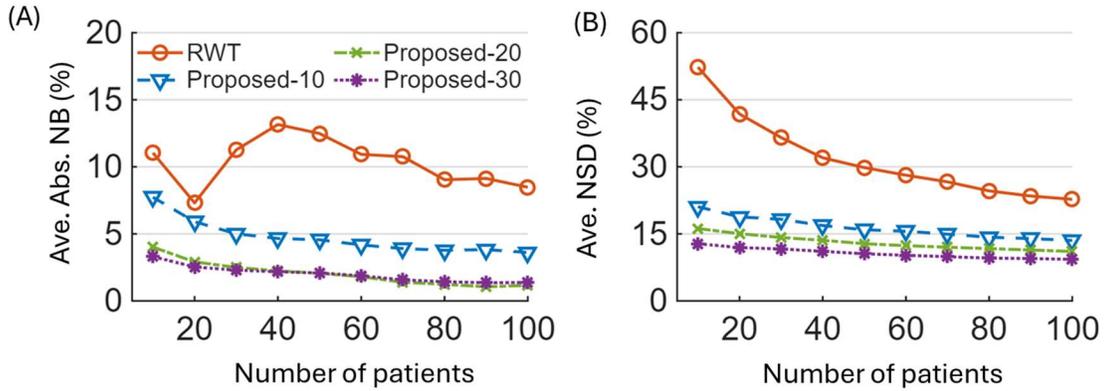

Figure 1. A comparison of RWT vs. the proposed approach with different known-ground-truth dataset sizes. In the legend, Proposed-X indicates the proposed approach with X number of ground-truth known samples. The results are presented for (A) average absolute normalized bias and (B) average normalized standard deviation in estimating NSR values in the numerical simulation studies.

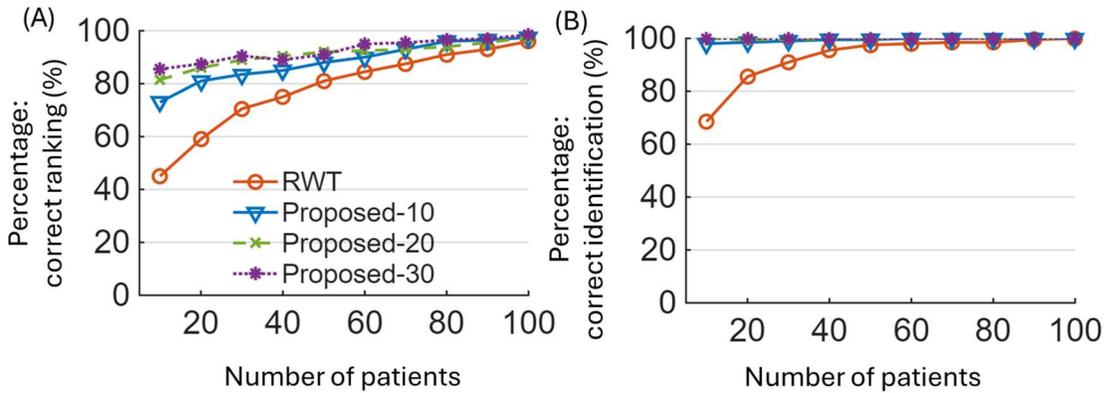

Figure 2. A comparison of RWT vs. the proposed approach with different known-ground-truth dataset sizes. In the legend, Proposed-X indicates the proposed approach with X number of ground-truth known samples. The results are presented for (A) correctly ranking all the QI methods and (B) correctly identifying the most precise method in the numerical simulation study.

The numerical studies conducted in this manuscript are in controlled settings where the assumptions made by the technique were satisfied. However, in clinically realistic scenarios, the assumptions may be violated. Thus, studying the performance of the method in clinically realistic scenarios is an important area of future research. These considerations motivate further validation of the proposed approach in clinically realistic settings. Another related area of future study is examining the performance of the proposed method in cases where the assumptions made by the technique are violated. These studies can also provide insights on further technical advances that are required to improve the performance of the method for more clinically realistic settings.

In conclusion, we extended regression without truth to integrate known-truth-data for evaluating quantitative imaging methods with limited patient data. Our results demonstrate that even small amounts of known-ground-truth data can substantially improve ranking performance. These findings suggest that the proposed approach may provide a practical mechanism for evaluating QI methods when patient data are limited and motivate further validation using clinical data.

## ACKNOWLEDGEMENTS

This work was supported by the National Institute of Biomedical Imaging and Bioengineering of the National Institute of Health under grants R01-EB031051, R01-EB031962 and NSF CAREER Award 2239707.